# A machine learning environment for evaluating autonomous driving software

Jussi Hanhirova, Anton Debner, Matias Hyyppä, Vesa Hirvisalo
Department of Computer Science, Aalto University, Finland
jussi.hanhirova@aalto.fi

*Abstract*—Autonomous vehicles need safe development and testing environments. Many traffic scenarios are such that they cannot be tested in the real world. We see hybrid photorealistic simulation as a viable tool for developing AI (artificial intelligence) software for autonomous driving. We present a machine learning environment for detecting autonomous vehicle corner case behavior. Our environment is based on connecting the CARLA simulation software to TensorFlow machine learning framework and custom AI client software. The AI client software receives data from a simulated world via virtual sensors and transforms the data into information using machine learning models. The AI clients control vehicles in the simulated world. Our environment monitors the state assumed by the vehicle AIs to the ground truth state derived from the simulation model. Our system can search for corner cases where the vehicle AI is unable to correctly understand the situation. In our paper, we present the overall hybrid simulator architecture and compare different configurations. We present performance measurements from real setups, and outline the main parameters affecting the hybrid simulator performance.

## I. INTRODUCTION

In this paper[1], we address learning environments for autonomous driving software subsystems that base their operation on machine learning. Typically, such subsystems are the parts of vehicle software systems that perceive the environment around the vehicle and initiate the automated reactions of the vehicle. The recent development of machine learning is enabling many previously human-operated tasks to be automated and allowing for many new applications, e.g., in the area of collaborative driving and smart mobility services.

The perception of the vehicle AI systems is usually based on sensors that yield video and LiDAR streams. The systems must process the sensor data related to environment perception in real-time, because they trigger actions with latency requirements. In the recent years, deep learning, particularly in the form of deep Convolutional Neural Networks (CNN), has become the dominant approach for implementing computer vision algorithms [1]. Because of the sharing needs of the applications and the performance properties of CNNs, there is a need for cloud service support and efficient hardware acceleration of the CNN inference computations. Novel hardware architectures and the emergence of Edge/Fog computing offer multiple options for further improvements of machine learning inference systems [2].

Training of machine learning models often calls for huge amounts of training data. Otherwise, the models might not be able to react accordingly in various situations that can arise while driving a vehicle. Especially demanding are exceptional driving situations, e.g., rare weather conditions, abnormal traffic conditions, and situations resulting in accidents. For such, the training itself and evaluation of the model behavior usually cannot be done in the real world. Typically, simulators are used instead [3,4,5].

Our research problem is to find the situations where a machine learning model does not yield the behavior that it should. Specifically, we try to search for corner cases of AI client software behavior. Corner case behavior usually means a situation that falls outside of the typical combination of operating parameters of a system. In the context of autonomous driving based on machine learning, this typically indicates that the machine learning is inadequate to cope with the driving situation.

Our contribution in this paper is to present a machine learning environment for detecting autonomous vehicle corner case behavior. Our system is based on connecting CARLA simulation software [6] to TensorFlow machine learning framework [7] and custom AI client software. The AI client software receives data from the model world via virtual sensors and transforms data into information using multiple CNN models. We evaluate multiple AI driven vehicles in the model world. The corner case detection mechanism relies on a ground truth from the simulation model, which makes detailed evaluation of distributed collaborative driving systems possible.

The contents and the structure of our paper is the following. We begin by describing our machine learning environment, in which our work has focused on ensuring sufficient performance as corner case search is time consuming. After

This work was supported by a Technology Industries of Finland Centennial Foundation grant

this, we describe our AI system, where we concentrate on evaluation and corner case search mechanisms. We continue our presentation by describing our experimentation on the hybrid simulator system performance. We end the paper with a discussion and our conclusions.

## II. MACHINE LEARNING ENVIRONMENT

Safety is a key requirement for autonomous driving systems. The traditional testing and validation methods try to guarantee that a system never enters unsafe states. Our machine learning environment uses state comparison to spot unexpected behavior. The environment cannot guarantee safe system behavior, but it can be used to test AI systems in challenging situations.

In this section, we describe our usage of the CARLA simulator and the TensorFlow framework as the basic building blocks of our environment. The next section describes our methodology for searching situations in which the AI client systems are not functioning as they should.

### A. CARLA

CARLA [6] is an open-source simulator for autonomous driving research built on top of the Unreal Engine[1]. CARLA consists of a simulation server and a client API. Using the API, a client can get sensor data from the simulated model world and control a vehicle.

We use CARLA as a part of our machine learning environment, which we use for evaluating autonomous driving software. Our system uses a custom corner case detector software, which is connected to CARLA to identify the situations where autonomous vehicle software under test is not working as supposed to. The corner case detector operates by observing and comparing the ground truth state of the CARLA simulation model and the state assumed by the connected CARLA clients.

### B. TensorFlow

We evaluate autonomous driving software that utilizes machine learning models to interpret sensor data from the simulated world. We use the TensorFlow [7] framework for training our machine learning models, and TensorFlow Serving[2] to deploy the models for inference.

### C. System overview

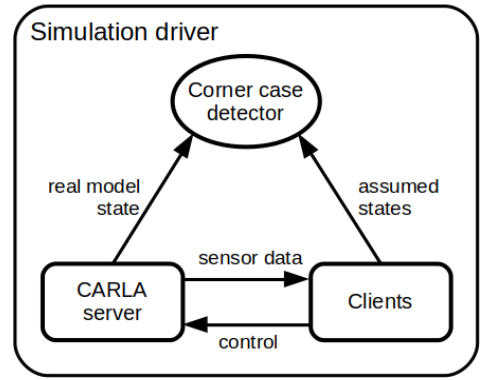

Fig. 1. The system main components are the corner case detector, the CARLA server, one or several AI clients and the simulation driver. The server and the clients form a simulation-control loop, and all of them pass state information to the corner case detector.

Our system consists of four main software components: a simulator driver, a corner case detector, the CARLA server and CARLA clients. An overview of the hybrid simulator system is presented in Figure 1. The simulation driver controls the system by starting and stopping the subsystem components based on the desired system configuration (e.g., the number and the type of vehicles, the CNN models).

The CARLA server executes the simulation model. Its main functions are rendering the client camera views to images and moving the client vehicles based on client control messages. CARLA clients are connected to the CARLA server, and they receive video frames from the simulated world and send vehicle control messages to the server.

In our system, the CARLA clients use a vehicle AI subsystem to autonomously control the vehicles. The AI subsystem receives sensor data from the client and uses CNN models to analyze the data. Based on CNN inference results, AI custom logic makes decisions on how to control the vehicle.

The corner case detector receives vehicle state information from both the CARLA server and the connected CARLA clients. The CARLA server provides the real vehicle state, while each CARLA client sends the currently observed and estimated state. The corner case detector compares the states, and if a defined condition is matched, then the simulation model state and the conflicting client observations can be saved for later analysis and AI model refinement.

## III. VEHICLE AI EVALUATION

Our approach allows evaluation of real AI software of varying complexity. The evaluation approach supports both software-based clients on virtual or shared platforms, and also AI software deployed on real vehicle AI hardware (hardware-in-the-loop).

---

[1] https://www.unrealengine.com
[2] https://www.tensorflow.org/serving/

Autonomous vehicle systems are usually very complex and require the use of special sensors and special hardware. In this paper, we present a simple artificial AI system that is implemented as fully software-based client that can be deployed on regular COTS platforms. System components can be encapsulated into containers and their deployment can be managed using tools such as Kubernetes[1].

*A. An example AI system*

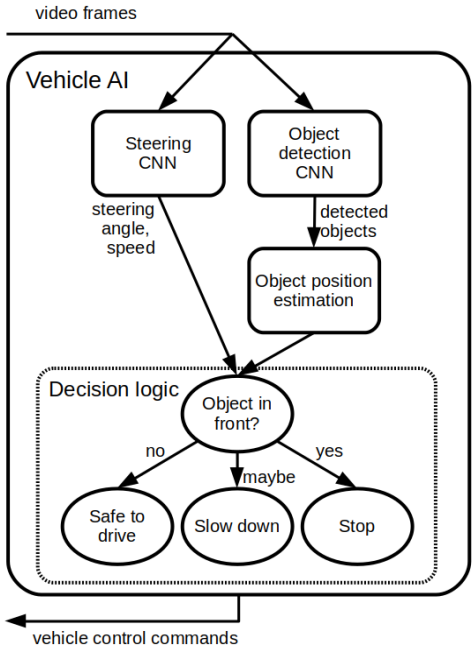

Fig. 2. Overview of the vehicle AI system.

Our example AI system has a basic functionality. It drives a vehicle along the simulation model roads and avoids collisions to objects by slowing down or stopping the vehicle. Overview of the vehicle AI system is presented in Figure 2. This vehicle AI system is used to extract information from sensor data (video frames) and to make control decisions based on the assumed vehicle state. It uses two CNNs to analyze the video data. An object position estimation subsystem and a decision logic subsystem are used to estimate the vehicle state. Finally, control commands are sent to the virtual vehicle in the simulation world.

The two CNNs used in the example AI system are an object detector CNN and a steering angle predictor CNN. The object detector CNN outputs the class of the detected object, a certainty for the detection, and bounding box information of the detected object location in the image. The steering CNN outputs the steering angle that needs to be applied to keep the vehicle on the road.

The object position estimation subsystem estimates the location of detected objects in the model world. Its operation is based on detecting the distance and location of objects in the front of the vehicle. This estimation is done by using the bounding box information from the object detection CNN. Based on the apparent size and the location of the bounding box on the image, the relative position of the detected object is estimated. Figure 3 presents how the estimation parameters are acquired.

Finally, the decision logic subsystem determines the vehicle state, and adjusts the speed or makes it to take a full stop. We consider three vehicle states: safe to drive, possibility for collision, and collision. When no objects are detected in front of the vehicle, the system steers the vehicle normally based on the output given by the steering CNN. If an object is close to the vehicle driving path, then driving speed is slowed down.

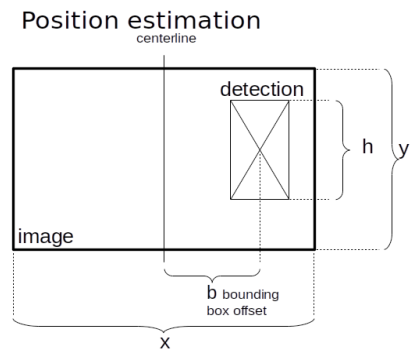

Fig. 3. Example of how the bounding box information can be used to estimate the relative position of an detected object. Comparing b to x and h to y, an estimate of the direction and distance from the camera location is computed.

*B. Modular AI composition*

We design AI systems by composing them from different machine learning models and custom control and reasoning software. Individual machine learning models that are aimed to perform a single task, such as to detect specific object classes or determine a suitable steering angle, are simpler to train and validate than, for example, models that aim to perform both of the tasks simultaneously. Similarly, re-training or fine-tuning individual machine learning models is faster to do than for more complex models.

From the AI system point-of-view, having different models for different tasks, and possible alternative models for same tasks, allows, for example, overriding uncertain model outputs with model control software. Similarly, changing entire machine learning models on-the-fly based on the observed system state is possible.

---

[1] https://kubernetes.io

## C. Corner Case Detection

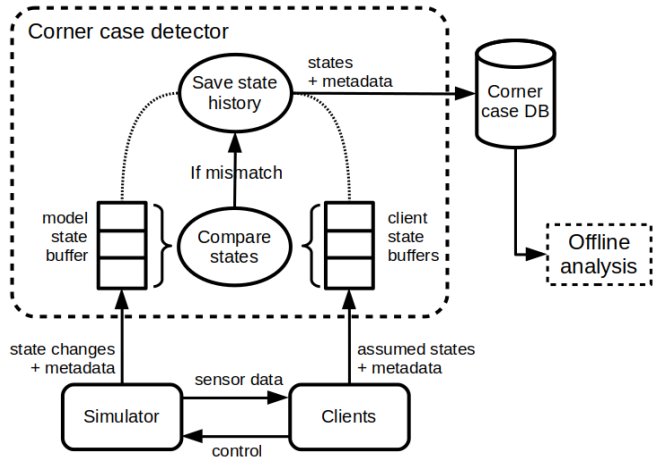

Fig. 4. The corner case detector receives state information from the simulator and from the connected clients. It detects state mismatches based on configured matching criteria. Situations leading to mismatches are saved to a database for later analysis.

The corner case detector searches for abnormal situations in the system. We use it to evaluate AI systems and to generate training data for machine learning. The search is based on comparing a combination of operating parameters. Upon finding a mismatch in the two states, both state histories can be saved for later offline analysis.

For example, in some scenario an AI client could assume that it is *safe to drive*, but the model state data can indicate that a collision has occurred. In this example, the corner case detection system would have found an abnormal situation.

### 1) State comparison

The corner case search subsystem receives state information from the simulator and from the connected AI clients. The simulator gives us *a ground truth* (the state of the world at a point in time) that we can compare against the AI system *estimate* (the perception of the world state at a time).

Vehicle AI states depend on the implementation of the AI system. Our example AI system (see Figure 2) has internal states for estimated object positions (object classes and their positions) and derived states that express the level of cautioness needed in driving. The derived states in the example AI are: *safe to drive*, *slow down*, *stop*.

If the AI system state estimate differs from the simulator ground truth then a mismatch is found. The criteria and tolerance for classifying state mismathing is simulation scenario specific. For example, in the case of comparing the estimated position of detected objects to their ground truth position, the mismatch tolerance can be the maximum position estimate error.

In our system, simulation model states and vehicle assumed states leading to the mismatch are saved to a database for later analysis. Temporary state buffers are used to hold a configurable amount of the state data history to allow reproduction of the situations leading to mismatching behavior.

### 2) Amount of data

Corner case search requires that the initial model state information is saved in the beginning of the search. During the search, only the state changes of the model need to be saved. In practise, the dynamic state updates are represented using three vectors (position, velocity, direction) of three 32-bit components (x,y,x). State updates are saved together with metadata composing of object identification information (32 bits) and timestamps (64 bits). All state changes can thus be expressed with 384 bits of data per moving simulation model entity.

The amount of required state information from client AIs depends on the AI implementation and also on the configuration of the corner case search. With the presented example AI system, the amount of produced state estimate is small when nothing is detected; the only thing to save is the state *safe to drive* with related meta-data. When something is detected then the amount of state information depends on the number of detected objects. Compared to the amount of data produced when rendering the camera views (320 x 240 x 3 x 8 bits / image [image resolution x color channels x color bit depth]), the amount of data produced in the corner case search is small.

### 3) Detecting timing anomalies

In real-time systems, timewise correct operation is a must. The corner case search can be used to detect misbehavior of AI logic reasoning related to computation timing. The AI client estimate of its state is based on sensor information that has been processed on multiple computational steps. By the time the state estimate is ready, the simulation model clock and the state have advanced. The state estimated by the client is thus always old when compared to the real state of the model.

The cyber-physical computer systems interacting with the physical world face many of the same challenges as in the simulator. Figure 5 illustrates the similarities. The physical environment can be directly observed, but the processing happens with software that is running on computer hardware. In addition to processing delays (the time interval from $t_0$ to $t_0\acute{}$), the use of software causes inherent inaccuracy in the understanding of timing as accessing time information (*i.e.*, physical clocks) cannot be done without delay. This can get dominant in distributed systems as communication delays are often longer than processing times.

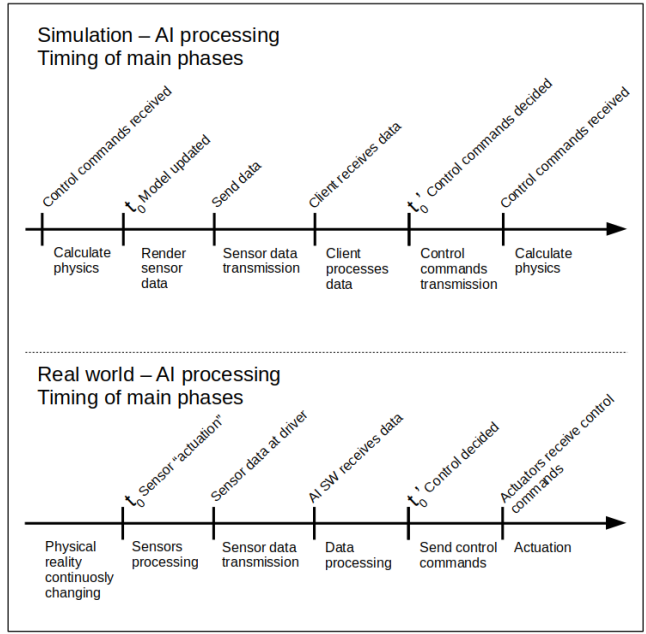

Fig. 5. AI systems perceive the environment by processing sensor data to information. Timing of the AI processing pipelines in real-world and in simulated setups have similarities in their main phases.

Our machine learning system is a hybrid one in the sense that it uses AI software running on real computer and communication hardware to process sensor data streams generated by distributed simulation software. On one hand, such an arrangement gives us very realistic understanding of the performance and the operation of the real-time AI. On the other hand, the arrangement calls for careful orchestration of the simulation system along the AI system.

Time related comparison is achieved by giving identifiers and timestamps to rendered camera frames so that the total time the frames spend on the system processing pipeline can be monitored. As we can use authentic software and hardware for the AI systems, we can observe the behavior caused by realistic latencies.

We can evaluate the effect of time delays on the behavior of AI systems by having the sensor data streams produced with a simulator. This is because we have full control and understanding of the state of the world that we perceive with AI. If we evaluate AI systems in the real physical world, then we are often restricted by our limited control and understanding of the ground truth.

## IV. SYSTEM CONFIGURATION

Our corner case search is fundamentally a random process, where real software is used to drive virtual vehicles in virtual model world. Search scenario specific criteria is configurable on the corner case detector subsystem. In order to make the search efficient, the system throughput needs to be maximized, so that virtual kilometers are gathered as efficiently as possible. At the same time, the system latencies need to be constrained so that the AI client software gets an up-to-date view of the virtual world.

The server-client communication in CARLA is asynchronous. Simulation physics advance independently without waiting for client input. Clients get sensor data from the server as fast as the server is capable to provide it. In practice, clients need an application-specific minimum number of frames per second to be able to realistically drive in the virtual world. A rough minimum for clients is 10 frames per second for basic driving, but higher frame rates are required to detect sudden situations.

In multi-client simulations and in scenarios with multiple interacting AI clients, each client requires a consistent view of the world within a tight-enough time frame. The jitter of the computation and the communication should not grow too dominating.

### A. Measurement setup

The simulator engine uses CPUs for physics calculation and GPUs for scene rendering. The presented AI client software uses GPU acceleration for CNN inference and CPU for the other operations. For efficient corner case search, sharing of the computational resources is required.

System components can be placed in several ways. The CARLA server, the AI clients and the CNN inference system can be placed on a shared host computer, or they can distributed on multiple hosts.

We study the system configuration alternatives using two heterogeneous host computers and by placing the system elements on them in three different ways. Figure 6 presents the three system configurations. The configuration A is a replicated system configuration and configurations B and C are distributed system configurations. In the replicated system (configuration A), all system elements are placed independently on the two host computers. In the distributed system (configurations B and C), the CNN inference system is placed on different host computer than the CARLA server and the AI clients.

We parametrize the system measurements by increasing the number of clients and measure the total system rendering throughput and the CNN model inference latencies and throughputs.

The measurement host 1 has an Intel i7-5820K CPU with 32GB RAM and a NVIDIA GeForce GTX 1050 TI GPU. The measurement host 2 is otherwise identical but it has a NVIDIA GeForce GTX 1080 TI GPU. We use CARLA version 0.9.0. The CNN model inference is done using TensorFlow Serving v1.8. The steering model CNN is a custom implementation of the model presented in [8]. The object detection model is the SSD Mobilenet v1 [9].

*B. Results*

Figures 7-13 present the measurement results from the system configurations presented in Figure 6. Figure 7 presents the total throughput of the two hosts in the replicated system configuration. In Figures 8 and 9 the latencies and per client throughputs of the same system configuration are presented. The latencies are on the left side vertical axis and the throughputs are on the right side vertical axis (the same way of presentation is also used in Figures 11 and 13). Using the replicated system configuration a maximum rendering capacity of 225 fps is reached with 6 vehicles. The per client rendering fps remains above 10 fps with up to 9 vehicles on host 1 and with up to 11 vehicles on host 2. The jitter of the object detection model grows rapidly after 9 vehicles on host 1 and after 5 vehicles on host 2. The steering model latency grows linearly and there is a slight increase in jitter after 4 vehicles with both hosts.

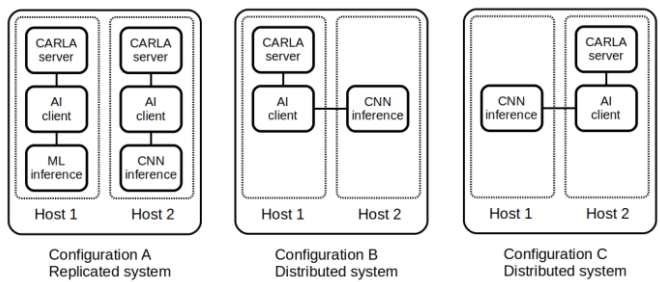

Fig. 6. System configurations used in experiments. Host 1 has a NVIDIA GeForce GTX 1050 TI GPU. Host 2 has a NVIDIA GeForce GTX 1080 TI GPU.

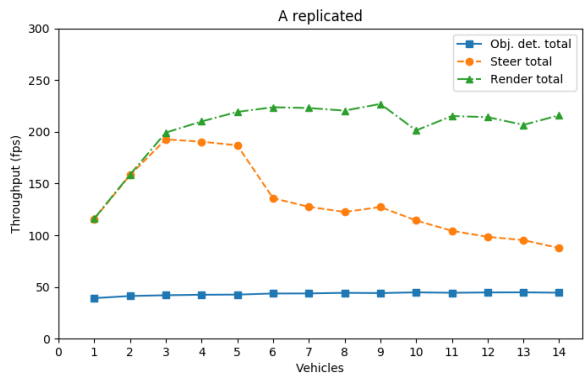

Fig. 7. Configuration A: Replicated system setup. All system components are placed on both of the host computers.

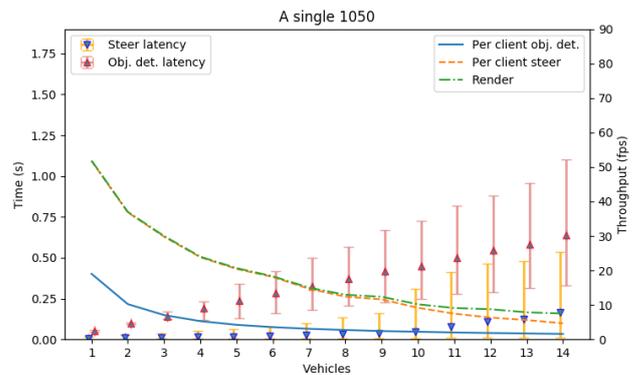

Fig. 8. Configuration A: Replicated system setup on host 1. All system components are placed on both of the host computers.

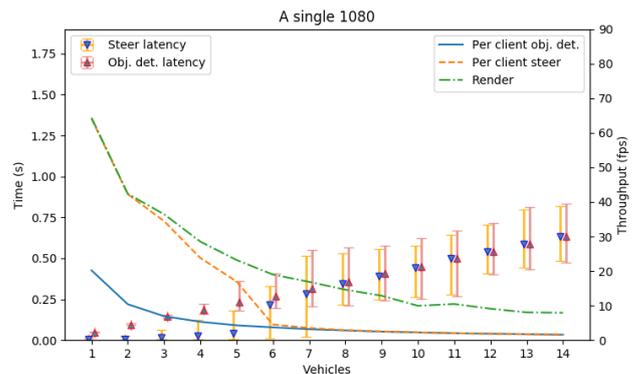

Fig. 9. Configuration A: Replicated system setup on host 2. All system components are placed on both of the host computers.

Figures 10 and 11 present results using the distributed system configuration B. Using this configuration maximum rendering capacity of 130 fps is reached with 5 vehicles (Figure 10). The per client fps remains above 10 fps with 12 vehicles (Figure 11). The jitter of the object detection model grows rapidly after 7 vehicles, also the mean inference time starts to grow after 9 vehicles. The steering model jitter starts to grow rapidly after 7 vehicles.

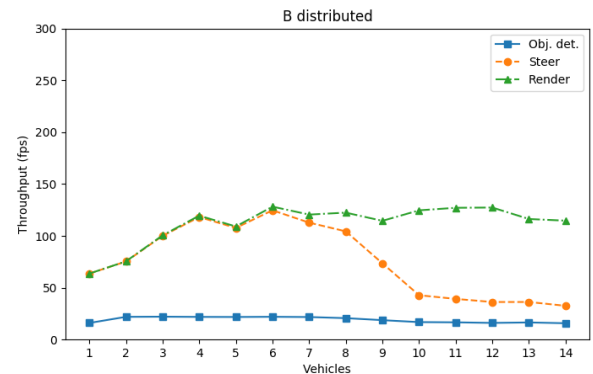

Fig. 10. Configuration B: Distributed system setup. System components are divided between the two host computers. CARLA server and the clients are placed on host 1 (1050 TI GPU) and the CNN inference system on host 2 (1080 TI GPU).

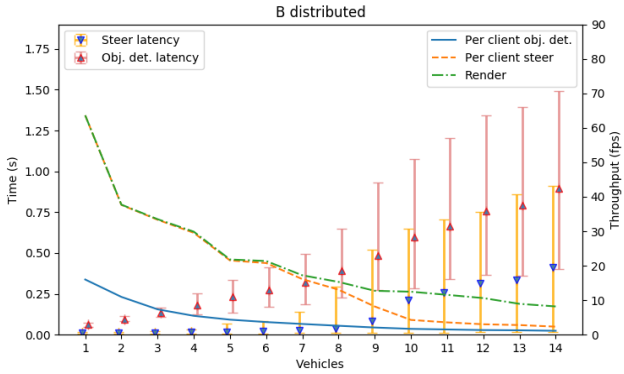

Fig. 11. Configuration B: Distributed system setup. System components are divided between the two host computers. CARLA server and the clients are placed on host 1 (1050 TI GPU) and the CNN inference system on host 2 (1080 TI GPU).

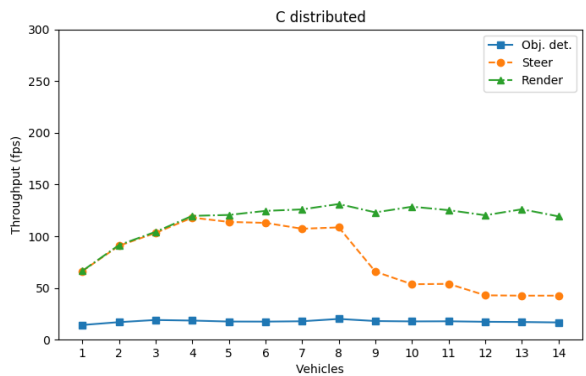

Fig. 12. Configuration C: Distributed system setup. System components are divided between the two host computers. CARLA server and the clients are placed on host 2 (1080 TI GPU) and the CNN inference system on host 1 (1050 TI GPU).

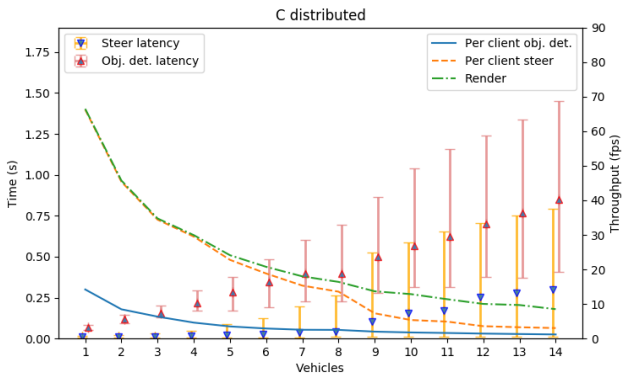

Fig. 13. Configuration C: Distributed system setup. System components are divided between the two host computers. CARLA server and the clients are placed on host 2 (1080 TI GPU) and the CNN inference system on host 1 (1050 TI GPU).

Figures 12 and 13 present the measurement results from the distributed system configuration C. Using this configuration maximum rendering capacity of 130 fps is reached with 7 vehicles (Figure 12). The per client fps remains above 10 fps with 12 vehicles (Figure 13). The jitter of the object detection model grows rapidly after 6 vehicles. The steering model latency and jitter grow almost linearly.

Maximum sustainable latency limits the number vehicles that can be simulated in multi-client configurations. Efficient search requires maximizing the number of vehicles so that virtual kilometers can be gathered as efficiently as possible. Optimal system configuration depends on many factors. For example, the details and requirements of the client AI software, the complexity of the simulation model and the quality of the rendered camera views define the maximum number of clients that can be simulated in a shared model in real-time.

The replicated setups do not have inter-host communication. But, there is complexity due to all system elements sharing the same computational resources. The performance behavior is linear on host 1 (1050 TI GPU) while on host 2 (1080 TI GPU) the steering model inference times saturate after 5 vehicles. The anomaly could be corrected by manual system runtime configurations.

## V. DISCUSSION

Hybrid simulation is an approach for autonomous driving and smart mobility research. It forms a safe machine learning environment for developing AI systems for challenging traffic situations. There are still a multitude of challenges with AI systems before fully autonomous systems could be deployed in the real world.

Main challenges from system implementation side are the computational and communicational latencies and how to predict them [10]. From AI systems point-of-view, the challenge is to develop methods for both robust and understandable AI models, which can act predictably in changing environments. Advances in communication technologies, such as 5G network slicing [11] for guaranteed bandwidth and QoS attempt to minimize the jitter related problems. Edge/Fog computing methods bring computational power close to clients, and thus minimize latency. Recent advances in machine learning (e.g., [12]) enlarge the boundaries of AI system capabilities.

Nevertheless, no model or simulation can capture the complexity of the physical reality. The unpredictable nature of reality and the arising abnormal situations make it impossible in practice to test all possible scenarios the AI systems might face. While the corner case search tries to cover as many as possible of these abnormal scenarios, it cannot guarantee the coverage of the search.

Configuration of the computational pipelines of the whole machine learning environment is vital for efficient search. Different search configurations call for different system configurations. For example, minimizing AI client computation latencies calls for different system configuration than maximizing the total throughput of the search. Due to system complexity it is hard to estimate actual system performance without direct measurements.

Configuration and management of the whole machine learning environment, composed of the simulator, AI clients and related subsystems, requires same tools and methods as the envisioned Edge/Fog systems [13]. While existing system orchestration and monitoring tools, such as Kubernetes or Mesos[1], can help in the management of computation, they also need to evolve with the Edge/Fog systems.

AI services deployed to the Edge/Fog datacenters are connected to moving AI clients in smart mobility scenarios, such as coordinated traffic management in intersections. This type of new services require mechanisms for, e.g., service discovery, communication protocols for AI system interoperability, dynamic load balancing and migration of computational state. Efficient system modeling and monitoring methods are needed to take into account the geographical distribution of the system elements and to adapt to dynamically moving clients and changing computational requirements.

Scaling the corner case search between multiple distributed hosts while keeping the simulation running in real-time faces the basic challenges of distributed computing. However, similar challenges are also faced in real vehicular systems as the software related to autonomous driving often must communicate, coordinate, sometimes also share resources with other systems present in the vehicles or their traffic environment. Hybrid simulation and corner case search connects to the wide field of smart mobility related research sought by academia and industry.

## VI. CONCLUSION

In this paper, we presented our machine learning environment and our experimental results on performance of the environment for autonomous driving software. We have used CARLA as the simulator software and TensorFlow as the AI engine in our hybrid simulator. Our measurements on the performance of the hybrid simulator show that sufficient performance for practical applications can be reached.

The technologies for autonomous driving systems are developing rapidly. Much of the recent development in AI systems is based on harnessing the computation power of novel computational accelerators. The basic design of such accelerators emphasizes their performance properties leaving their real-time properties complicated. The advent of Edge/Fog computing makes new options available, but it also further complicates the computation needed in AI applications. The development work of AI systems often concentrates on normal and simple behavior leaving abnormal and complicated behavior uncovered. We see hybrid photorealistic simulation as one of the tools to overcome such challenges.

REFERENCES

[1] Yann LeCun, Yoshua Bengio, Geoffrey, Hinton. (2015). Deep Learning. Nature. 521. 436-44. doi: 10.1038/nature14539.
[2] Xiaowei Xu, et al, Scaling for edge inference of deep neural networks. Nature Electronics, Vol 1, pp. 216-222, April 2018. https://vast.cs.ucla.edu/publications/scaling-edge-inference-deep-neural-networks
[3] NVIDIA DRIVE Constellation simulator. https://www.nvidia.com/en-us/self-driving-cars/drive-constellation/
[4] AImotive. aiSim2. https://aimotive.com/aisim2/
[5] Mario Garzón, Anne Spalanzani. An hybrid simulation tool for autonomous cars in very high traffic scenarios. *ICARCV 2018*. https://hal.inria.fr/hal-01872
[6] Alexey Dosovitskiy, German Ros, Felipe Codevilla, Antonio Lopez, and Vladlen Koltun. 2017. CARLA: An Open Urban Driving Simulator. In Proceedings of the 1st Annual Conference on Robot Learning.
[7] Martín Abadi et al. 2015. TensorFlow: Large-Scale Machine Learning on Heterogeneous Systems. 2015. https://www.tensorflow.org/
[8] Mariusz Bojarski et al. 2016. End to End Learning for Self-Driving Cars. 2016. https://arxiv.org/abs/1604.07316
[9] Andrew G. Howard, et al. 2017. MobileNets: Efficient Convolutional Neural Networks for Mobile Vision Applications. http://arxiv.org/abs/1704.04861
[10] Jussi Hanhirova, Vesa Hirvisalo, Anton Debner and Matias Hyyppä. AI Accelerator Latencies In Hybrid Vehicular Simulation. 2019. http://workshops.inf.ed.ac.uk/edla/papers/2019/EDLA2019_paper_7.pdf
[11] Ibrahim Afolabi, Tarik Taleb, Konstantinos Samdanis, Adlen Ksentini and Hannu Flinck. Network Slicing and Softwarization: A Survey on Principles, Enabling Technologies, and Solutions. in *IEEE Communications Surveys & Tutorials*, vol. 20, no. 3, pp. 2429-2453, thirdquarter 2018. doi: 10.1109/COMST.2018.2815638
[12] Isabeau Prémont-Schwarz, Alexander Ilin, Tele Hao, Antti Rasmus, Rinu Boney, Harri Valpola. Recurrent Ladder Networks. NIPS 2017. http://papers.nips.cc/paper/7182-recurrent-ladder-networks.pdf
[13] Yuan Ai, Mugen Peng, Kecheng Zhang. 2018. Edge computing technologies for Internet of Things: a primer. Digital Communications and Networks. doi: 10.1016/j.dcan.2017.07.001.

---

[1] mesos.apache.org